\documentclass[12pt]{article}
\usepackage{graphicx}
\usepackage{booktabs} 
\usepackage{upgreek}
\usepackage{xcolor}
\usepackage{caption}
\usepackage{subcaption}
\usepackage{cite}
\usepackage{tikz}

\usepackage{hyperref}  
\hypersetup{
    colorlinks=true,       
    linkcolor=black,          
    citecolor=black,        
    filecolor=black,     
    urlcolor=black          
}

\def\pbnr{}
\def\speaker{William Fawcett\footnote{Contact \texttt{william.fawcett@cern.ch}}  and Shanzhen Chen}
\def\onbehalfof{the LHCb Collaboration}
\def\title{Can LHCb Study Three Body Decays with Neutrals?}
\def\affiliation{School of Physics and Astronomy\\
The University of Manchester, Manchester, UK}

\newcommand\pubnumber{\pbnr}
\newcommand\pubdate{\today}
\def\Title#1{\begin{center} {\Large #1 } \end{center}}
\def\Author#1{\begin{center}{ \sc #1} \end{center}}

\newcommand{\OnBehalf}[1]{\sbox0{#1}\ifdim\wd0=0pt
        {}
	\else
	{\\on behalf of #1}
	\fi}
\newcommand{\SupportedBy}[1]{\sbox0{#1}\ifdim\wd0=0pt
        {}
	\else
	{\footnote{#1}}
	\fi}
\def\Address#1{\begin{center}{ \it #1} \end{center}}

\newcommand\pubblock{\includegraphics[width=5cm]{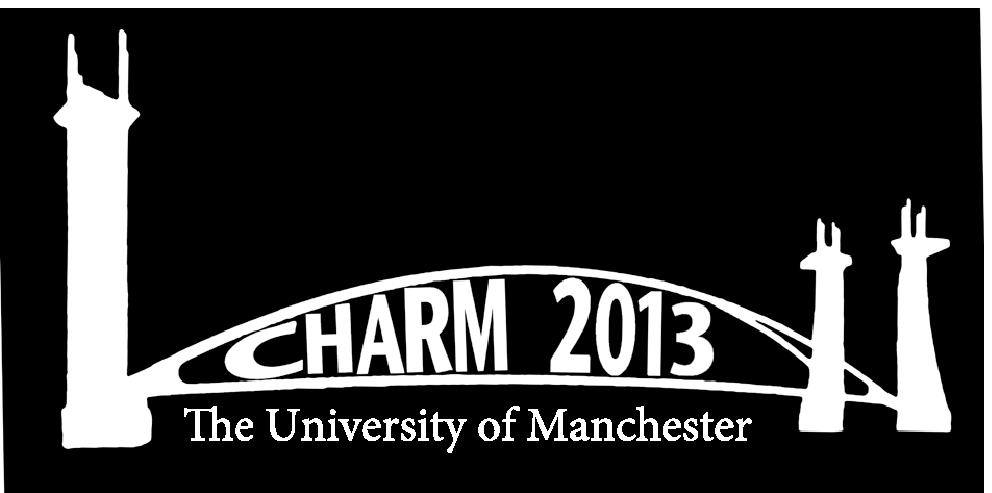}\hfill{\begin{tabular}{l} \pubnumber\\
         \pubdate  \end{tabular}}}
\newenvironment{Abstract}{\begin{quotation}  }{\end{quotation}}
\newenvironment{Presented}{\begin{quotation} \begin{center} 
             PRESENTED AT\end{center}\bigskip 
      \begin{center}\begin{large}}{\end{large}\end{center} \end{quotation}}

\def\venue{The 6$^{th}$ International Workshop on Charm Physics\\
(CHARM 2013)\\
Manchester, UK,  31 August -- 4 September, 2013}

\def\beq{\begin{equation}}
\def\eeq#1{\label{#1}\end{equation}}
\def\eeqn{\end{equation}}

\def\beqa{\begin{eqnarray}}
\def\eeqa#1{\label{#1}\end{eqnarray}}
\def\eeqan{\end{eqnarray}}

\let\bar=\overbar

\def\Dslash{\not{\hbox{\kern-4pt $D$}}}
\def\dslash{\not{\hbox{\kern-2pt $\del$}}}

\def\msb{{\bar{\ssstyle M \kern -1pt S}}}

\begin{document}
\begin{titlepage}
\pubblock

\vfill
\Title{\title}
\vfill
\Author{\speaker\OnBehalf{\onbehalfof}}
\Address{\affiliation}
\vfill
\begin{Abstract}
We present the first attempt to use a new method to measure CP violation in Dalitz plots. This method is unbinned, model independent and has a greater sensitivity to CP violating effects than binned methods. Preliminary studies have been made using the three-body decays $D^0 \rightarrow K_\textup{s}^0 h^+ h^-$ and $D^0 \rightarrow h^+ h^- \pi^0$, which are especially challenging since there is one neutral particle in each of the final states. An attempt to visualise where CP violation occurs in Dalitz plots is also presented.

\end{Abstract}
\vfill
\begin{Presented}
\venue
\end{Presented}
\vfill
\end{titlepage}
\def\thefootnote{\fnsymbol{footnote}}
\setcounter{footnote}{0}

\section{Introduction}

LHCb is attempting to study three-body decays with neutral particles in the final state.  Several decay modes of $D^0$ mesons were selected for analysis: $D^0 \rightarrow K_{\textup{s}} h^+ h^-$ and $D^0 \rightarrow h^+ h^- \pi^0$ where $h=K$ or $\pi$ (note that all decays imply their charge conjugate modes). This is the first time these modes have been studied at LHCb, though searches for asymmetries in these modes have been carried out at previous experiments \cite{Aaltonen:2012nd,Zupanc:2009sy,delAmoSanchez:2010xz,Kopp:2000gv,Asner:2003uz}. Due to the large $c \bar{c}$ and $b\bar{b}$ cross-sections \cite{Aaij:2013mga,Aaij:2013noa}, LHCb will collect large quantities of charm decays \cite{Bediaga:2012py} allowing for improvements in these measurements. In 2011, M.~Williams presented a new method of measuring CP violation in multi-body decays \cite{Williams:2011cd}. The method is unbinned, model-independent, and was shown to have increased sensitivity to CP violating effects relative to binned methods \cite{Williams:2011cd}. In this document we present the first attempt to use this method.

\section{The LHCb Detector}
The LHCb detector \cite{Alves:2008zz} is a single-arm forward spectrometer covering the pseudorapidity range $2 < \eta < 5$, designed for the study of particles containing $b$ or $c$ quarks. The detector includes a high precision tracking system consisting of a silicon-strip vertex detector surrounding the $pp$ interaction region (VELO), a large-area silicon-strip detector located upstream of a dipole magnet with a bending power of about 4\,Tm, and three stations of silicon-strip detectors and straw drift tubes placed downstream. The combined tracking system has a momentum resolution $\Delta p/p$ that varies from 0.4\% at 5\,GeV/$c$ to 0.6\% at 100\,GeV/$c$, and an impact parameter resolution of 20\,$\upmu$m for tracks with high transverse momentum. Charged hadrons are identified using two ring-imaging Cherenkov detectors. Photon, electron and hadron candidates are identified by a calorimeter system consisting of scintillating-pad and preshower detectors, an electromagnetic calorimeter and a hadronic calorimeter. Muons are identified by a system composed of alternating layers of iron and multiwire proportional chambers. The trigger \cite{Sjostrand:2006za} consists of a hardware stage, based on information from the calorimeter and muon systems, followed by a software stage which applies a full event reconstruction.

\section{Preliminary Selection}
Firstly it is important to test whether or not the modes of interest can be selected with sufficient purity at LHCb. Therefore a preliminary selection has been performed on the modes $D^0 \rightarrow K_\textup{s} K^+ K^-$, $D^0 \rightarrow K_\textup{s} \pi^+ \pi^-$, $D^0 \rightarrow K^- \pi^+ \pi^0$  and $D^0 \rightarrow \pi^+ \pi^- \pi^0$. A series of rectangular cuts were applied to kinematic variables, and  Boosted Decision Trees were used to create a multi-variate discriminator to further reduce background. 

In this note we consider $D^0$ mesons that are produced in the semi-leptonic decays of $B$ mesons: $\overline{B} \rightarrow D^0 \mu^- \bar{\nu} (X)$, where $X$ indicates that additional decay products may be present. The flavour of the $D^0$ is identified by the charge of the muon, and so this sample is referred to as single-tagged. A subset of these decays are of the form:  $\overline{B} \rightarrow D^{*+} \mu^- \bar{\nu} (X), D^{*+} \rightarrow D^0 \pi_{s}^{+}$. In this case the $D^0$ flavour is identified by both the charge muon and the charge of the slow pion ($\pi_s^+$), so this sample is referred to as double-tagged. Preliminary yields for single- and double-tagged samples are shown in Table \ref{tab:STresults} and Table \ref{tab:DTresults} respectively. The selection and yield measurements were carried out separately for $K_\textup{s}$ mesons reconstructed inside and outside the LHCb VELO acceptance, but the results have been combined for brevity. Note that purity, $P$, has been defined as
\begin{equation}
P=\frac{S}{S+B}
\end{equation}
where $S$ and $B$ are the number of signal and background candidates (within a selected $D^0$ mass region), respectively. 

\begin{table}
\begin{subtable}{.5\linewidth}\centering
{\begin{tabular}{l |c c}
\textbf{Mode}  & \textbf{Yield} & \textbf{Purity} (\%) \\ \hline
$K_\textup{s} \pi^+ \pi^-$ & $935{,}000$ & 84 \\
$K_\textup{s} K^+ K^-$ & $116{,}000$ & $81$ \\ \hline
\end{tabular}}
\caption{Single-tagged.}\label{tab:STresults}
\end{subtable}%
\begin{subtable}{.5\linewidth}\centering
{\begin{tabular}{l |c c}
\textbf{Mode} & \textbf{Yield} & \textbf{Purity} (\%) \\ \hline
$K_\textup{s} \pi^+ \pi^-$ & $182{,}000$	& $91$ \\
$K_\textup{s} K^+ K^-$ & $25{,}000$ & 91 \\ \hline
\end{tabular}}
\caption{Double-tagged.}\label{tab:DTresults}
\end{subtable}
\caption{Preliminary yields and purities for $D^0\rightarrow K_\textup{s} h^+ h^-$ modes sourced from semileptonic $B$ decays. These yields correspond to an integrated luminosity of 2\,fb$^{-1}$, collected at $\sqrt{s} = 8$\,TeV by the LHCb spectrometer in 2012. Note that the combined dataset form 2011 and 2012 corresponds to an integrated luminosity of 3\,fb$^{-1}$.}\label{tab:1}
\end{table}

\section{Unbinned CP Violation Search}
Typically, model-independent measurements of CP violation in three body decays are done using binned Dalitz plots. It is only the average result of each bin that is compared, not each event, hence sensitivity to CP violation is reduced or lost entirely if bins cover parts of the Dalitz plot with differing CP asymmetries. We therefore adopt the unbinned method proposed in Ref.~\cite{Williams:2011cd}. It is a two-sample test on the data for $Y \rightarrow abc$ and charge conjugate decays and involves computation of the following test statistic;
\begin{equation}
T=  \frac{1}{n(n-1)} \sum_{i,j>i}^n \psi (\Delta {\bf{x}}_{ij}) + \frac{1}{\bar{n}(\bar{n}-1)} \sum_{i,j>i}^{\bar{n}} \psi (\Delta {\bf{x}}_{ij}) - \frac{1}{n\bar{n}} \sum_{i,j}^{n,\bar{n}} \psi (\Delta {\bf{x}}_{ij}),
\label{eq:T}
\end{equation}
where  $n$ $(\bar{n}$) is the number of $Y$ ($\bar{Y}$) decays, $\Delta {\bf{x}}_{ij}=|{\bf{x}}_i - {\bf{x}}_j|$ is the distance between events $i$ and $j$ in Dalitz space with ${\bf{x}}\equiv (m_{ab}^2,m_{bc}^2,m^2_{ac})$ and $m_{ab}$ is the invariant mass of the two-body system $ab$. The function $ \psi (\Delta {\bf{x}}_{ij})$ is a weighting function that diminishes with increasing distance. This ensures that only points close to one another contribute significantly to $T$. Hence, $T$ compares the average distances of all particle and antiparticle events among each other with the average distance of particle to antiparticle events and so tests for consistency between the two distributions. The function chosen for this analysis was
\begin{equation} 
 \psi (\Delta {\bf{x}}_{ij}) = - \frac{6(k\Delta {\bf{x}}_{ij} -1)}{k{ \Delta \bf{x}}_{ij}+4\sqrt{k{\Delta \bf{x}}_{ij}}+1}
\end{equation}
with $k=0.2$. This function was not the original proposed in Ref.~\cite{Williams:2011cd}, but is a simplification made in order to reduce computation time.

In Equation \ref{eq:T}, the first sum runs over all particles, the second sum runs over all antiparticles and the third term sums over the interaction of the two. In the no-CP violation hypothesis, $T=0$. This implies that any deviation from this would be evidence for CP violation. In order to determine whether a deviation is statistically significant, we use a permutation method described below.

\subsection{Permutation Method}
A control sample with no CP violation can be obtained from the signal data sample by randomly relabeling each event as originating from either a $D^0$ or $\overline{D}^0$. This is repeated many times (in the case of this analysis, 100 times) in order to generate a series of pseudo-data sets. From each of these pseudo-data sets, $T$ is then recalculated. The $p$-value is then defined as the percentage of trials where the new $T$ value is greater than the original.

\begin{figure}[htb]
\centering
\includegraphics[height=2.5in]{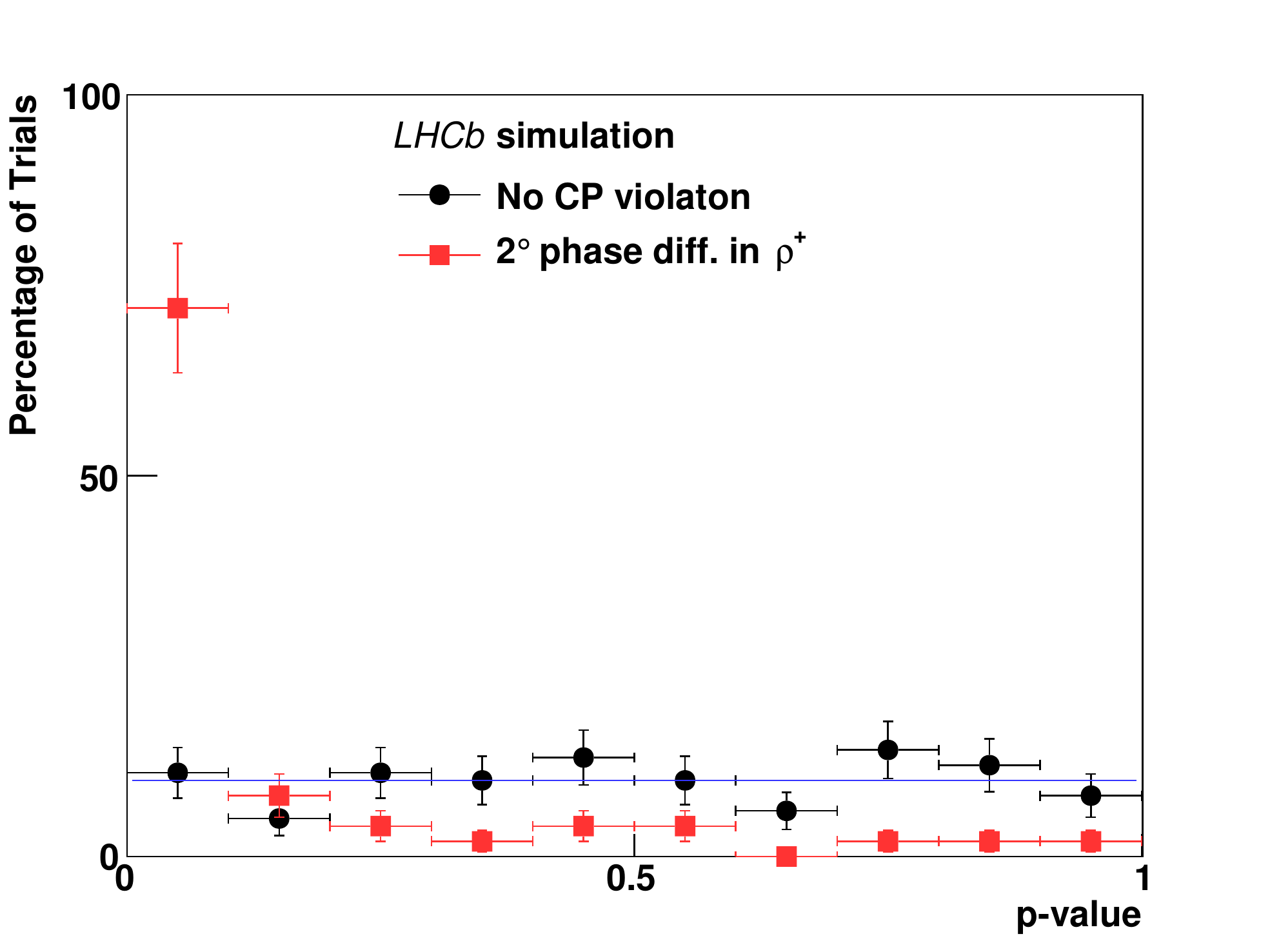}
\caption{$p$-value distributions for toy MC samples either without CP violation (black circles) or with CP violation (red squares). The horizontal blue line shows the no-CP violation hypothesis. 72\% of samples with CP violation had $p<0.1$ in this example.}
\label{fig:pvals}
\end{figure}

In order to determine the sensitivity of the method to CP asymmetries, many toy Monte Carlo (MC) data sets are created, some with and some without CP violation. The permutation method is run for each sample and the $p$-value is calculated each time. We can then compare the distribution of $p$-values for samples with and without CP violation. An example of such a comparison is displayed in Figure \ref{fig:pvals}. This demonstrates that the method is sensitive to CP-violating asymmetries and, under the assumption of a particular model, will allow us to quantify that sensitivity.

\subsection{Equilateral Dalitz Plots}
Dalitz plots usually have two axes which are combinations of invariant mass squared of two of the daughter particles. For example, one could choose $m^2_{ab}$ vs.~$m^2_{bc}$ from the decay $Y\rightarrow abc$. One problem with calculating $T$ is that the value of the distance between points, $\Delta {\bf{x}}_{ij}$ depends on the choice of axes used. This therefore introduces an arbitrary dependence of $T$ upon the combination of axes chosen. One solution would be to use all three axes and create a 3D Dalitz plot. This would remove the choice of axis combinations altogether, however this would increase the already significant computational time required to calculate $T$ as three coordinates would be required.

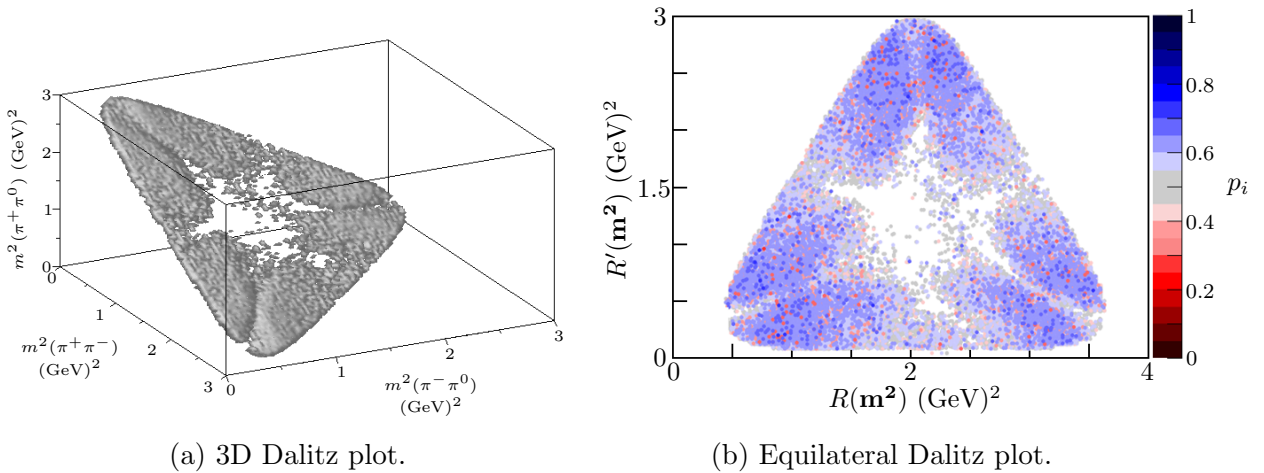
\begin{figure} 
        \centering
\hspace{-20pt}
        \begin{subfigure}[b]{0.5\textwidth}
               \centering
\begin{tikzpicture} [>=latex]
              \node[anchor=south west,inner sep=0] at (0,0) {\includegraphics[width=\textwidth]{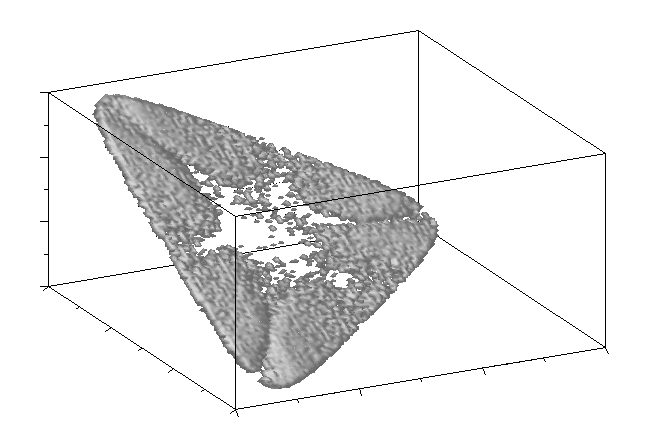}};

\tiny

\node at (0.39,1.76) {0};
\node at (0.41, 2.52) {1};
\node at (0.39, 3.28) {2};
\node at (0.39, 4.05) {3};

\node at (0.46,1.62) {0};
\node at (1.1,1.12) {1};
\node at (1.79, 0.66) {2};
\node at (2.54, 0.17) {3};

\node at (2.84,0.13) {0};
\node at (4.31,0.36) {1};
\node at (5.75,0.6) {2};
\node at (7.2,0.85) {3};

\node at (0, 2.8) [rotate =90] {$m^2(\pi^+ \pi^0)$ (GeV)$^2$};
\node at (0.7, 0.7) {$m^2(\pi^+ \pi^-)$};
\node at (0.7,0.4) {(GeV)$^2$};
\node at (5.5, 0.2) {$m^2(\pi^- \pi^0)$};
\node at (5.5,-0.1) {(GeV)$^2$};

\end{tikzpicture}
                \caption{3D Dalitz plot.}
               \label{fig:simulation}
        \end{subfigure}%
        ~ 
        \begin{subfigure}[b]{0.5\textwidth}
               \centering
\begin{tikzpicture} [>=latex]
              \node[anchor=south west,inner sep=0] at (0,0) {\includegraphics[width=\textwidth]{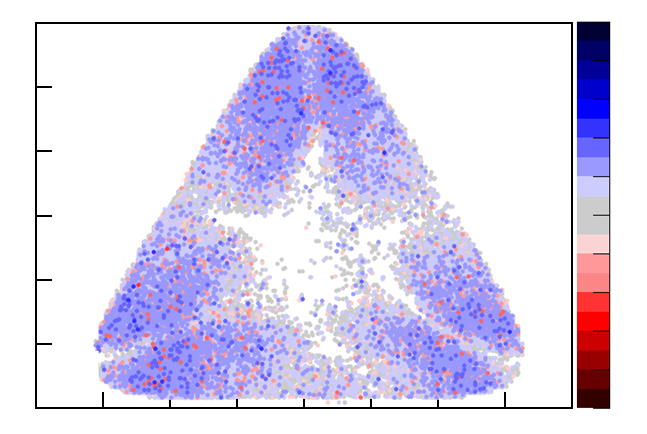}};
\footnotesize

\node at (0.26,0.32) {0};
\node at (0.18,2.57) {1.5};
\node at (0.26,4.85) {3};

\node at (0.44,0.13) {0};
\node at (3.59,0.13) {2};
\node at (6.74,0.13) {4};

\node at (7.95,2.59) {$p_i$};

\scriptsize
\node at (7.33,0.3) {0};
\node at (7.44,1.2) {0.2};
\node at (7.44,2.12) {0.4};
\node at (7.44,3.03) {0.6};
\node at (7.44,3.94) {0.8};
\node at (7.33,4.85) {1};

\footnotesize
\node at (3.65,-0.25) {$R({\bf{m^2}}$) (GeV)$^2$};
\node at (-0.3,2.55) [rotate=90] {$R'({\bf{m^2}})$ (GeV)$^2$};

\end{tikzpicture}
               \caption{Equilateral Dalitz plot.}
                \label{fig:eqdal}
        \end{subfigure}
        ~ 
      \caption{Dalitz plots containing simulated $D^0/\overline{D}^0 \rightarrow \pi^+ \pi^- \pi^0$ decays with no CP violation. Traditional Dalitz plots are projections of (a) onto two of the axes shown. Note that in (b) the $p_i$ colour scale is defined in Section~\ref{sec:viz} and has been included for comparison later.} \label{fig:Dalitz}
\end{figure}

Our solution has been to create an \emph{equilateral} Dalitz plot. This is a projection of a plane $m_{ab}^2 + m_{bc}^2 + m_{ca}^2 = Constant$ onto a 2D surface. An example of the result is shown in Figure \ref{fig:Dalitz}. This provides advantages over traditional Dalitz plots, as any dependence of $T$ on axis choice has been removed and only two coordinates are required to calculate  $\Delta {\bf{x}}_{ij}$. The disadvantages are that the abscissa and ordinate axes have lost any obvious physical meaning. We label the axes in equilateral Dalitz plot with $R({\bf m ^2})$ and $R'({\bf m^2})$, which denote the appropriate mapping of points in 3D Dalitz space onto two axes.

\subsection{Visualisation}
\label{sec:viz}
As proposed in Ref.~\cite{Williams:2011cd} we can calculate the contribution of each event in a Dalitz plot to $T$ as follows
\begin{equation}
T_i = \frac{1}{2n(n-1)} \sum_{j \neq i}^n \psi(\Delta {\bf{x}}_{ij}) - \frac{1}{n\bar{n}} \sum_j^{\bar{n}} \psi(\Delta {\bf{x}}_{ij}),
\end{equation}
where the first sum runs over particles and the second runs over antiparticles. A similar equation is also defined for the charge conjugate data set. We can then run a permutation test and generate a $p_i$ value, defined as the proportion of permutation trials with a $T_i$ value greater than the original. In the case of no CP violation, we expect $p_i$ values to be close to 0.5. In the case of CP violation, $p_i$ values tend towards either 0 or 1. This allows us to colour code each event with its $p_i$ value and display this on a Dalitz plot, giving us a visual indication of the areas in which CP violation is occurring. As with the $p$-value, the $p_i$ value does not indicate the magnitude of CP violation, only the likelihood of it occurring.

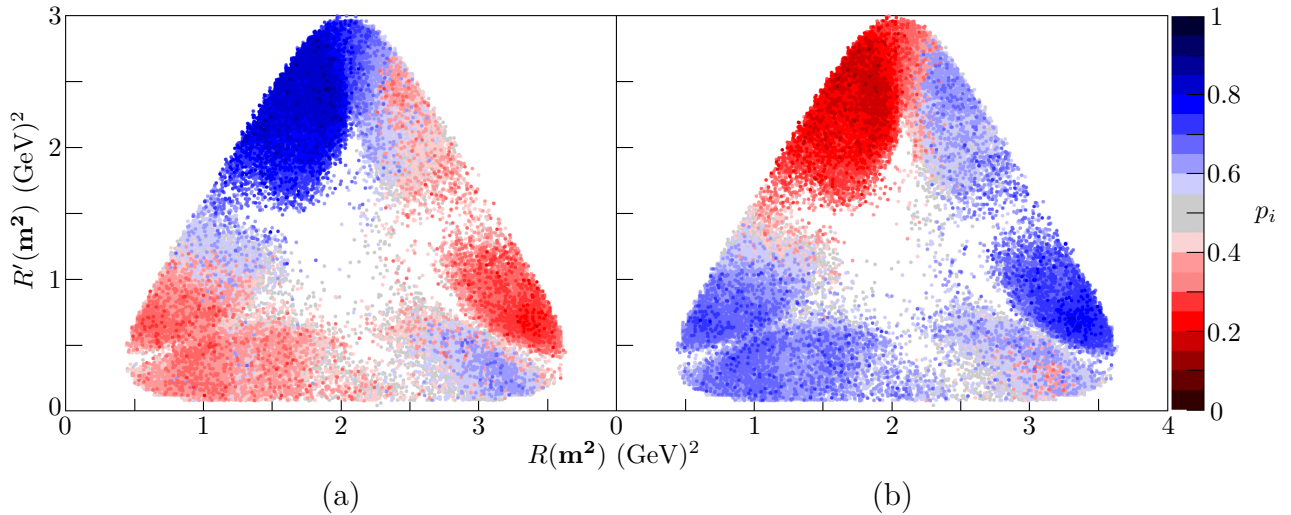
\begin{figure}[htb]

\begin{tikzpicture} [>=latex]
              \node[anchor=south west,inner sep=0] at (-1.5,0) {\includegraphics[width=\linewidth]{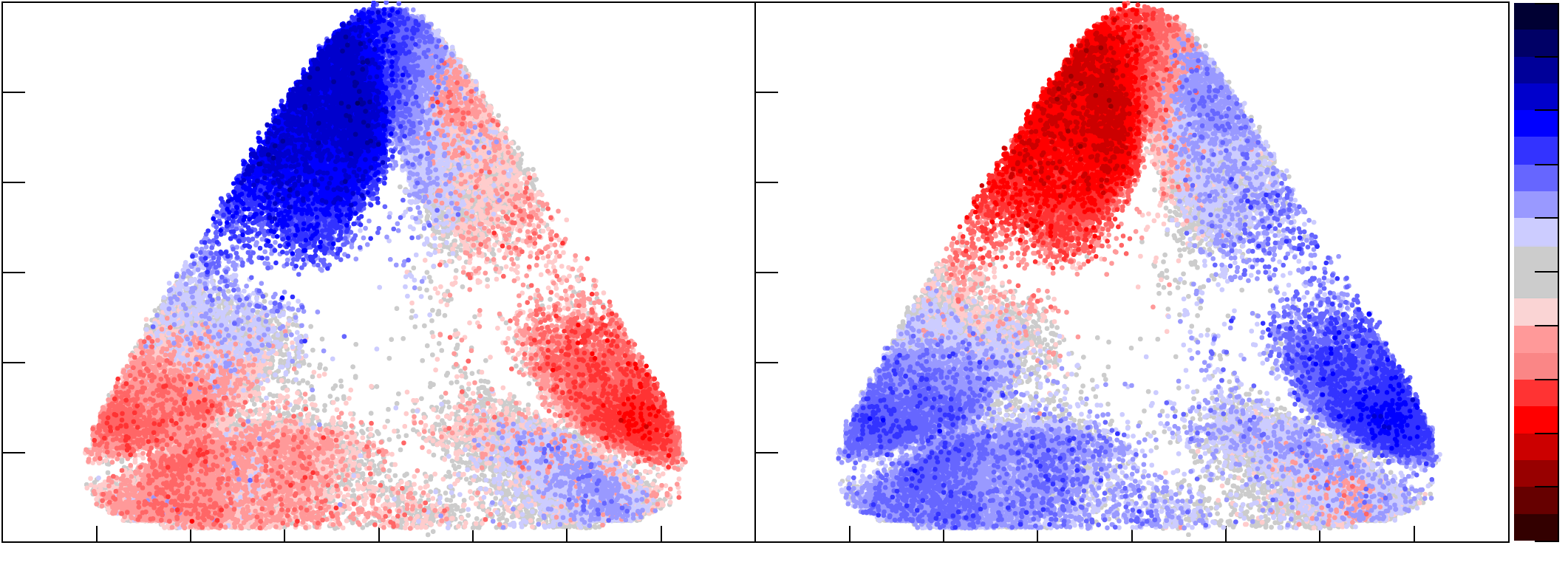}};
\hspace{-43pt}
\footnotesize

\node at (-0.1,0.45) {0};
\node at (-0.1,2.14) {1};
\node at (-0.1,3.9) {2};
\node at (-0.1,5.65) {3};

\node at (0.03,0.2) {0};
\node at (1.87,0.2) {1};
\node at (3.7,0.2) {2};
\node at (5.53,0.2) {3};

\node at (7.36,0.2) {0};
\node at (9.20,0.2) {1};
\node at (11.03,0.2) {2};
\node at (12.85,0.2) {3};
\node at (14.7, 0.2) {4};

\node at (15.34,0.39) {0};
\node at (15.44,1.44) {0.2};
\node at (15.44,2.5) {0.4};
\node at (15.44,3.55) {0.6};
\node at (15.44,4.6) {0.8};
\node at (15.34, 5.64) {1};

\footnotesize{
\node at (16, 3) {$p_i$};

\footnotesize
\node at (7.33,-0.2) {$R({\bf{m^2}}$) (GeV)$^2$};
\node at (-0.5,3.2) [rotate=90] {$R'({\bf{m^2}})$ (GeV)$^2$};

\normalsize
\node at (3.7,-0.75) {(a)};
\node at (11.03,-0.75) {(b)};

}

\end{tikzpicture}
\caption{Each of the above is an equilateral Dalitz plot showing the $p_i$ values of simulated $D^0/ \overline{D}^0 \rightarrow \pi^+ \pi^- \pi^0$ decays with $1^\circ$ of phase difference in the $\rho(770)^+$ between $D^0$ and $\overline{D}^0$. In the left (right) plot we have chosen to plot \emph{only} $\overline{D}^0$ ($D^0$) decays.}
\label{fig:CPV}
\end{figure}

Figure \ref{fig:CPV} shows plots containing simulated $D^0 / \overline{D}^0 \rightarrow \pi^+ \pi^- \pi^0$ decays, generated by a toy MC program. In Figure \ref{fig:CPV}a only $\overline{D}^0$ decays are shown whereas \ref{fig:CPV}b shows only $D^0$ decays (note that we need both $D^0$ and $\overline{D^0}$ to calculate the $p_i$ value for each point, however we can choose to plot the two separately). Both \ref{fig:CPV}a and \ref{fig:CPV}b show darker red and blue areas, indicating regions where the $p_i$ value has differed from the no-CP violation case. However the reader will also notice that the colours are inverted on the two plots -- this is as one would expect as each plot shows the antiparticle of the other. Either of these figures should be compared to Fig. \ref{fig:eqdal} which shows the same decays except with no CP violation.

\section{Summary}
We have shown that the the development of a new method of measuring CP violation in Dalitz plots. At this time a measurement can only permits us to test for the presence of a CP asymmetry, rather than measure it quantitatively. We have also presented the development of a method of visualising where CP violation occurs in a Dalitz plane. In addition we have transformed the traditional 2D Dalitz plot into an equilateral Dalitz plot which has several advantages when computing $T$ including reduced computation time.

\end{document}